\documentclass[a4paper,twocolumn,superscriptaddress,prl,floatfix]{revtex4}
\usepackage{graphicx}          
\usepackage{dcolumn}
\usepackage{amssymb,amsmath}
\usepackage{color}

\begin{document}
\title{Laser structuring for control of coupling between THz light and phonon modes}
\author{X. W. Wang}
\affiliation{Centre for Micro-Photonics, Faculty of Engineering
and Industrial Sciences, Swinburne University of Technology,
Hawthorn, VIC 3122, Australia}

\author{G. Seniutinas}
\affiliation{Centre for Micro-Photonics, Faculty of Engineering
and Industrial Sciences, Swinburne University of Technology,
Hawthorn, VIC 3122, Australia} \affiliation{School of Mathematical
and Physical Sciences, University of Technology Sydney, Thomas St,
Ultimo, NSW 2007 Australia}

\author{A. Bal\v{c}ytis}
\affiliation{Centre for Micro-Photonics, Faculty of Engineering
and Industrial Sciences, Swinburne University of Technology,
Hawthorn, VIC 3122, Australia}\affiliation{Center for Physical
Sciences and Technology, Saul\.{e}tekio al. 3, LT-10222 Vilnius,
Lithuania}

\author{I. Ka\v{s}alynas}
\affiliation{Center for Physical Sciences and Technology,
Saul\.{e}tekio al. 3, LT-10222 Vilnius, Lithuania}

\author{V. Jak\v{s}tas}
\affiliation{Center for Physical Sciences and Technology,
Saul\.{e}tekio al. 3, LT-10222 Vilnius, Lithuania}

\author{V. Janonis}
\affiliation{Center for Physical Sciences and Technology,
Saul\.{e}tekio al. 3, LT-10222 Vilnius, Lithuania}

\author{R. Venckevi\v{c}ius}
\affiliation{Center for Physical Sciences and Technology,
Saul\.{e}tekio al. 3, LT-10222 Vilnius, Lithuania}

\author{R. Buividas}
\affiliation{Centre for Micro-Photonics, Faculty of Engineering
and Industrial Sciences, Swinburne University of Technology,
Hawthorn, VIC 3122, Australia}

\author{D. Appadoo}
\affiliation{Australian Synchrotron, Blackburn Road, Clayton,
Victoria 3168, Australia}

\author{G. Valu\v{s}is}
\affiliation{Center for Physical Sciences and Technology,
Saul\.{e}tekio al. 3, LT-10222 Vilnius, Lithuania}

\author{S. Juodkazis}
\affiliation{Centre for Micro-Photonics, Faculty of Engineering
and Industrial Sciences, Swinburne University of Technology,
Hawthorn, VIC 3122, Australia}

\date{\today}
\begin{abstract}
Modification of surface and volume of sapphire is shown to affect
reflected and transmitted light at THz spectral range. Structural
modifications were made using ultra-short 230~fs laser pulses at
1030 and 257.5~nm wavelengths forming surface ripples of $\sim
250$~nm and 60~nm period, respectively. Softening of the
transverse optical phonon TO$_1$ mode due to disorder was the most
pronounced in reflection from laser ablated surface. It shown that
sub-surface periodic patterns of laser damage sites have also
modified reflection spectrum due to coupling of THz radiation with
phonons. Application potential of laser structuring and
disordering for phononic engineering is discussed.
\end{abstract}
\maketitle
\section{Introduction}

Spectral properties at sub-1~mm wavelengths at around terahertz
(1~THz = $10^{12}$Hz) frequencies are important for understanding
weak interaction in peptides and proteins~\cite{Mlnonl,Png},
material response at vicinity of phase
transitions~\cite{Doucet,Barnes}, glass formation where low
frequency Raman spectra are showing the low frequency
10-50~cm$^{-1}$ boson peak due to rearrangement of density of
states in amorphous materials~\cite{Piekarz}. In silk, amorphous
and crystalline structural components with proteins forming a 3D
network of random and $\alpha$-coils together with a crystalline
$\beta$-sheet phase can be distinguished in THz spectral
window~\cite{Liu,Kim,15n18299,Sun,Hu1,15a11863}.

Control of phonon spectrum is a new frontier in material science
for growth and deposition of layered structures of usually
incompatible materials with different thermal expansion
coefficients. Materials' optical properties at IR range can be
engineered using control of phonon spectrum. For example, surface
phonon polaritons (SPP) are shown to be in control of
directionality of black body emission when coupled with surface
gratings~\cite{Greffet}. By introducing patterns with period
$\Lambda$ satisfying grating equation:
\begin{equation}\label{e1}
   \frac{2\pi}{\lambda}\sin\theta = k_{\parallel} +
m\frac{2\pi}{\Lambda},
\end{equation}
\noindent where $\lambda$ is the wavelength of emitted light
(black body radiation), $\theta$ is the angle of emitted light,
$k_{\parallel}$ is the surface wave component parallel to the
surface (SPP), and $m$ is the integer. It is possible to create a
strong directional outcoupling of the IR emission into free space
out of the sample~\cite{Greffet}. With a surface grating $\Lambda
= 0.55\lambda$ directional emission was observed at $\lambda =
11.36~\mu$m from SiC surface with intensity 20 times larger as
from a flat surface at the same temperature~\cite{Greffet}. This
is called Wolf's effect: at the angle, $\theta$, of the largest
emissivity, the reflectivity, $R$, has a dip. This corresponds to
Kirchhoff's law: the polarized directional spectral emissivity
$\varepsilon$ is equal to the absorptivity $\alpha$ and is given
by $\varepsilon \equiv \alpha = 1 - R$.

Surface nanotexturing of sapphire by fs-laser ablation is shown to
enhance light extraction efficiency in light emitting
diodes~\cite{15jpd285104}. Surface patterns with periods about
1-5~$\mu$m with different duty cycle are used for epitaxial
lateral overgrowth (ELO) a GaN to reduce density of threading
dislocations. A better understanding of nanotexturing effects at
T-ray spectral range (phonon modes) is required for future
material engineering. Nano-/micro-scale surface modifications can
be readily made by direct writing with ultra-short fs-laser
ablation.

\begin{figure}[bt]
\begin{center}
\includegraphics[width=8.5cm]{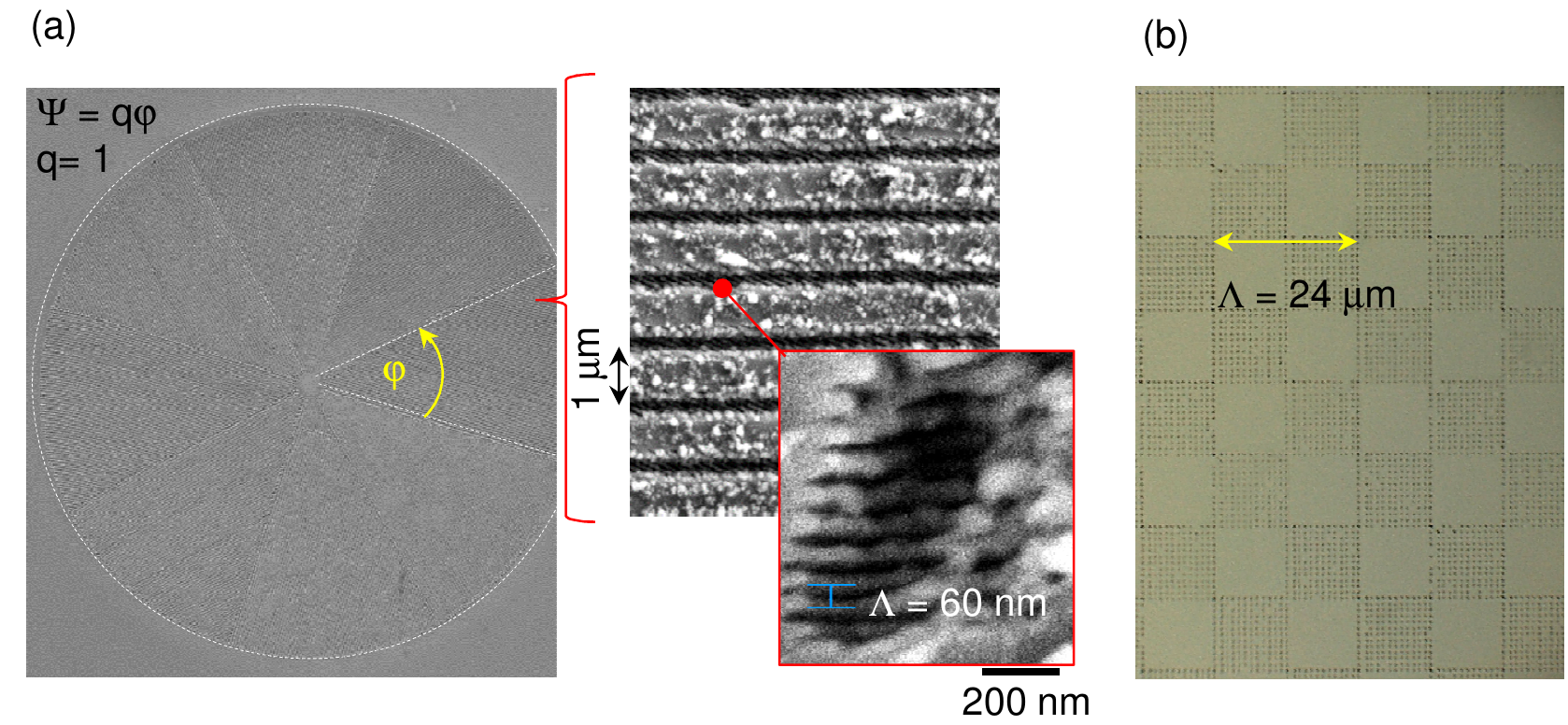}
\caption{(a) SEM images of a $q$-plate pattern recorded on
sapphire at different magnifications. The azimuthal dependence of
the slow axis is given by $\Psi = q\varphi$ where $q = 1$ and
$\varphi$ is the polar angle. Conditions: pulse energy $E_p =
4.25$~nJ (on sample), $\lambda = 257$~nm wavelength, $\tau_p =
230$~fs pulse duration, $v = 1$~mm/s scan speed, repetition rate
$f = 200$~kHz, focusing with $NA = 0.4$ objective lens (irradiance
$I_p = 3.8$~TW/cm$^2$/pulse). Diameter of $q$-plate is 200~$\mu$m
and fabrication time 4~min. (b) Photo image of a grating inscribed
in sapphire at 20-$\mu$m-depth by fs-laser 1030~nm/230~fs pulses
using numerical aperture $NA = 1.42$ objective lens at $v =
1$~mm/s, $f = 50$~kHz, $E_p = 1.52~\mu$J/pulse.} \label{f-samp}
\end{center}
\end{figure}

Here, laser modifications on/in sapphire are recognised in
reflection at THz spectral range. Such nano-/micro-scale patterns
can be used for controlling phonon spectrum on the surface and in
the bulk of polar semiconductors and dielectrics, hence, affecting
heat transport, emissivity, coupling mechanisms between surface
excitation in layered structures~\cite{Ishitani}.

\begin{figure}[bt]
\begin{center}
\includegraphics[width=6.5cm]{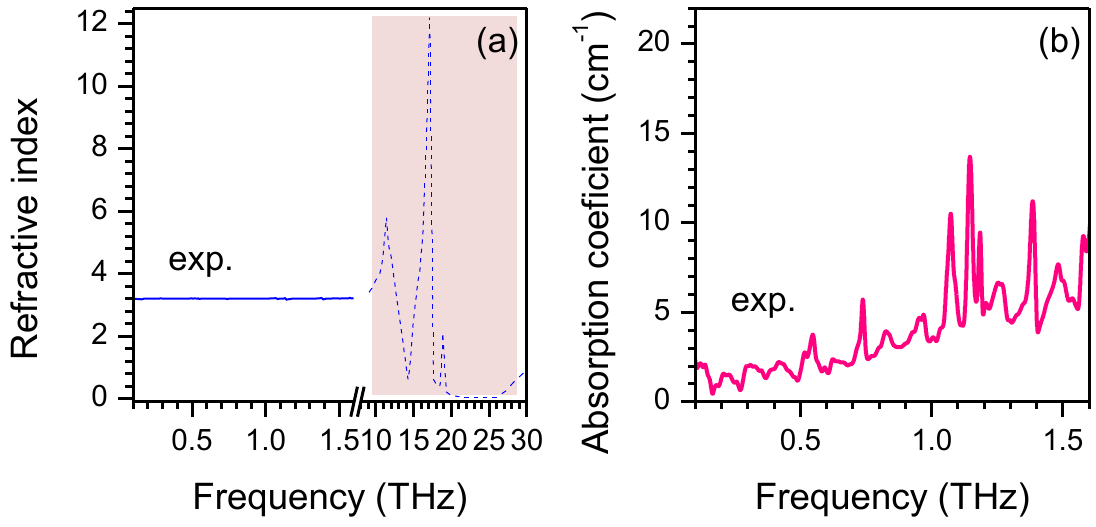}
\caption{Measured (\emph{exp.}) refractive index, $n$, (a) and
absorption coefficient, $\alpha_s$, (b) of sapphire (c-face) in
T-ray region. Tabulated $n$ values~\cite{Dobrovinskaya} for higher
THz range are added in (a). The refractive index and absorption
coefficient of sample were measured with a  THz-TDS system
(Teravil-Ekspla). } \label{f-sapp}
\end{center}
\end{figure}

\begin{figure}[bt]
\begin{center}
\includegraphics[width=8.5cm]{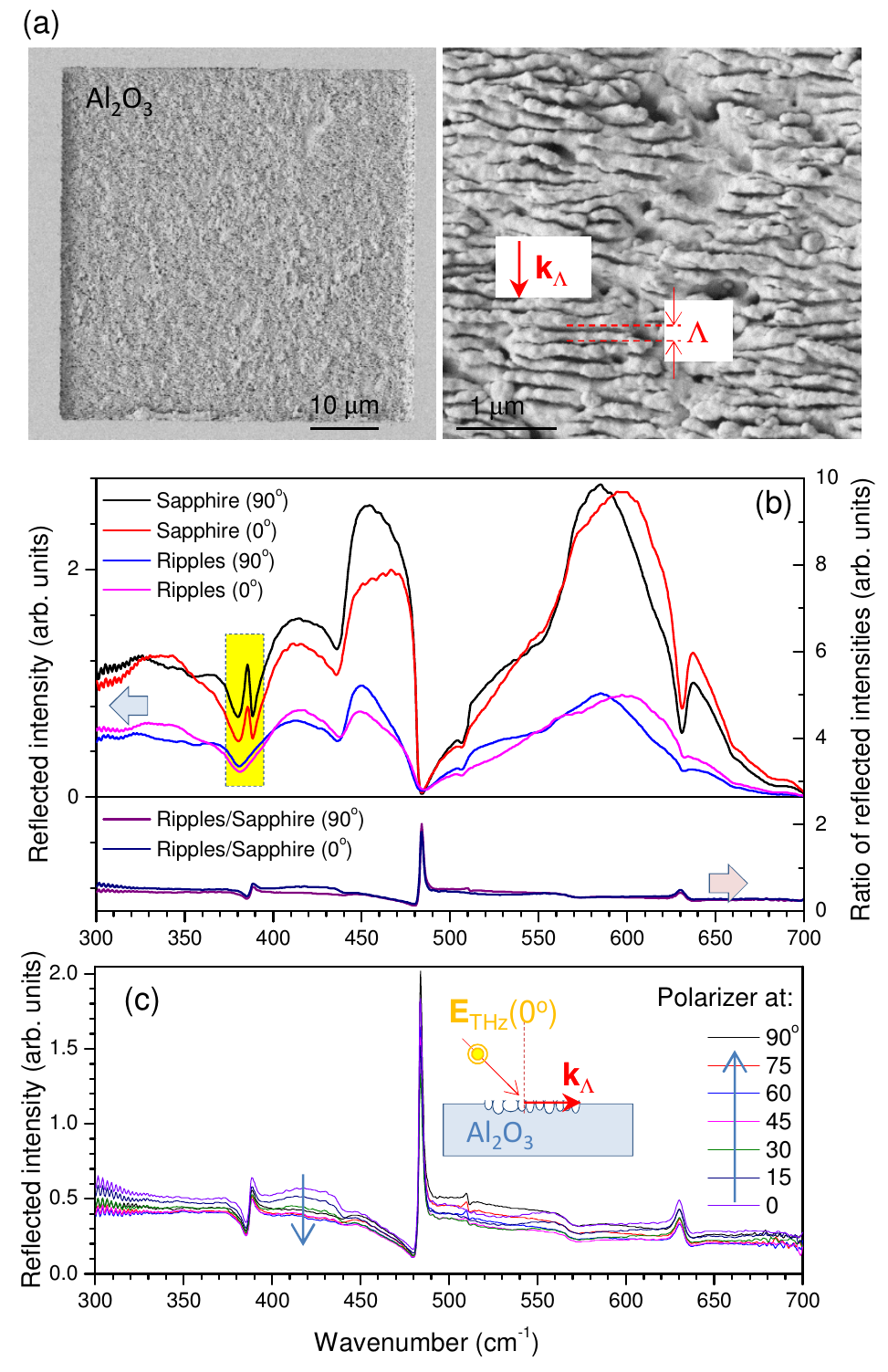}
\caption{(a) Typical SEM images of ripples laser ablated at
1030~nm/280~fs at different magnifications. Period of ripples
$\Lambda$ and $\mathbf{k}_\Lambda$ is the wavevector.  (b)
Reflection from optical sapphire flat and ablated surface
(ripples) at two polarisations $0^\circ$ and $90^\circ$; a metal
wire grating polariser was used with $0^\circ$-orientation
corresponding to polarisation ``aligned'' (inset) with the surface
ripples (or $\mathbf{E_{\mathrm{THz}}}\perp \mathbf{k}_\Lambda$;
$k_\Lambda = 2\pi/\Lambda$); angle of incidence is close to
normal. The highlighted box region shows spectral location of the
strongest modifications at the sapphire's TO$_1$ phonon mode
$385$~cm$^{-1}$. (c) Normalised reflectivity at several
polarisation angles. Inset shows geometry of experiment. T-ray
source was IR beamline at the Australian synchrotron.}
\label{f-ripples}
\end{center}
\end{figure}

\begin{figure}[tb]
\begin{center}
\includegraphics[width=8.5cm]{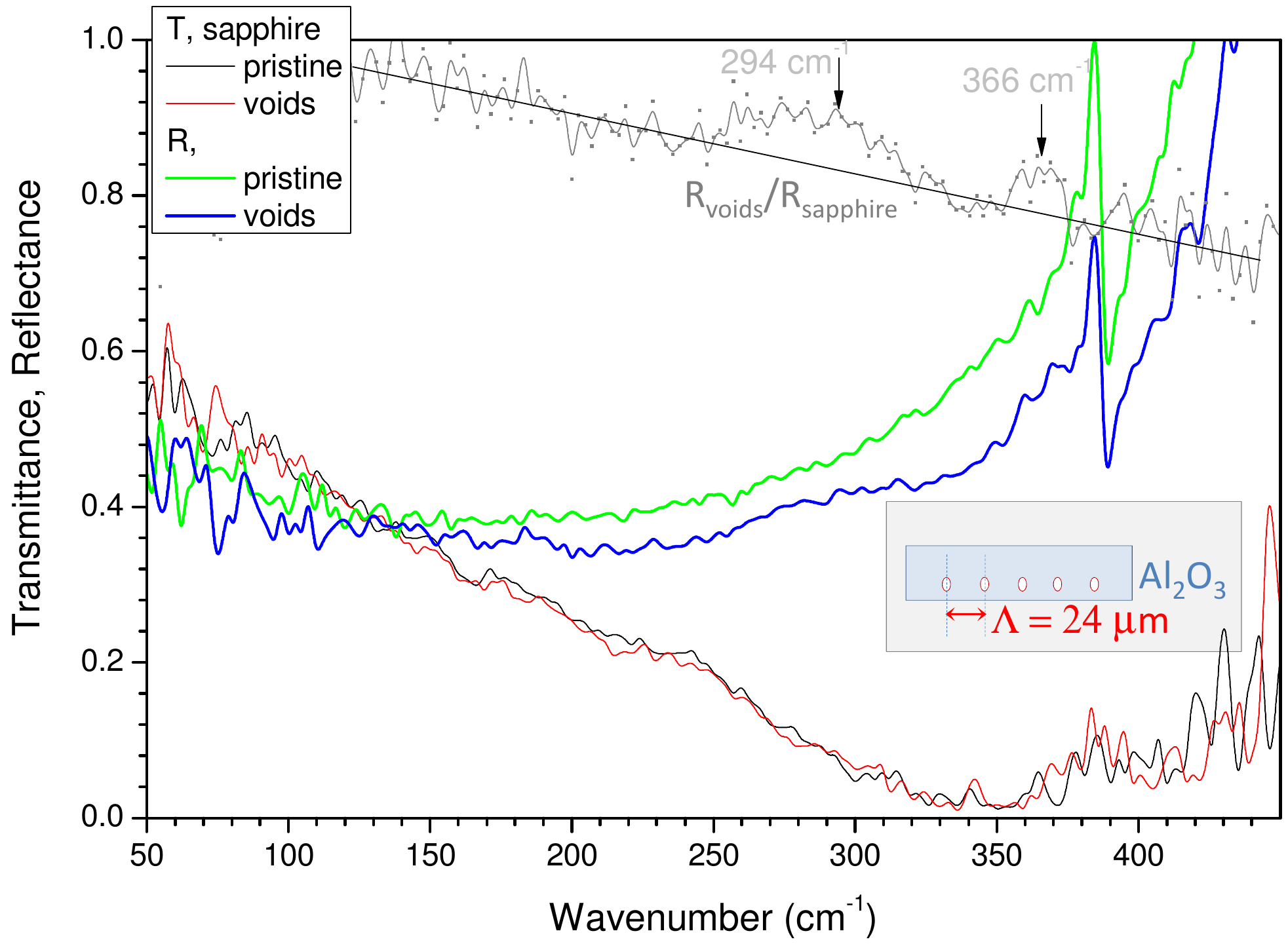}
\caption{Reflection and transmission spectra at normal incidence
from 365-$\mu$m-thick two-side polished sapphire flats without and
with void grating inscribed at $20~\mu$m depth. At wavenumbers
larger than 400~cm$^{-1}$ there are artifacts due to due to a low
efficiency of the beam splitter used. Inset shows schematically a
side view of the sample.} \label{f-ref}
\end{center}
\end{figure}

\section{Samples and procedures}

The THz/Far-IR Beamline at the Australian synchrotron was used to
characterise laser modified samples at 40 - 800~cm$^{-1}$ spectral
range. The beamline is equipped with a Bruker IFS 125/HR Fourier
Transform (FT) spectrometer and Opus software was used for initial
data analysis. Up to 100 spectral scans were captured and
averaged to improve signal-to-noise (S/N) ratio. All measurements were at room temperature. 
Also, the spectra of reflection and transmission  at normal
incidence were obtained for the range of 20-450~cm$^{-1}$ with a
resolution of 4~cm$^{-1}$, using a customized vacuum
Fourier-transform infrared (FT-IR) spectrometer and averaging up
to 100 spectral scans.

Samples of c-plane sapphire (Shinkosha Ltd.) were used for
femtosecond (fs-)laser structuring at $\lambda = 1030$ and
257.5~nm wavelengths (Pharos, Light Conversion). Fabrication
conditions at  $\lambda = 257$~nm: pulse duration $\tau_p \sim
230$~fs, repetition rate of $f = 0.2$~MHz, linear scan speed of $v
= 1$~mm/s. Exposure pattern was controlled via an integrated
software-hardware solution (Workshop of Photonics, Ltd) equipped
with Aerotech stages. Focusing objective lens of numerical
aperture $NA = 0.4$ (50$^\times$ magnification, PlanApo UV,
Mitutoyo), which focused into focal spot with waist (radius) $w_0
= 0.61\lambda/NA \simeq 390$~nm. The number of pulses per focal
spot was $N = 2w_0\times f/v = 157$. Fabrication conditions at
$\lambda = 1030$~nm were: $NA = 1.42$ (100$^\times$ magnification,
PlanApo NIR, Mitutoyo) at $v = 1$~mm/s, $f = 50$~kHz, $E_p =
1.52~\mu$J/pulse. A single pulse irradiance would correspond to
$I_p = 1.1$~PW/cm$^2$ and pulse power of $P_p = 6.6$~MW, which is
above the self-focusing threshold. Together with a spherical
aberration due to a deep focal spot position this smears axially
the pulse energy, however, it is above the intrinsic threshold of
void formation $\sim 10$~TW/cm$^2$~\cite{04pra042903}. In both
cases, on the surface and in the bulk a strong structural damage
was induced.

\section{Results and discussion}

Figure~\ref{f-samp}(a) shows a pattern of grooves comprising an
inhomogeneous birefringent plate with a topological defect of
charge $q$ (hence $q$-plate) for the azimuthal patterning of the
optical axis. The azimuthal dependence of the slow axis is given
by $\Psi = q\varphi$, where, $q$ is the half integer and $\varphi$
is the polar angle. The laser ablated regions have nanoscale
pattern of ripples with period scaling with wavelength and the
refractive index, $n$, as $\Lambda = \lambda/(2n)$ on dielectric
(transparent materials) surfaces. Hence, by choosing wavelength
different nano-structures can be formed. At tight focusing below
the surface, voids can be created inside dielectric host material
when a single pulse irradiance is $I_p > 10$~TW/cm$^2$
(Fig.~\ref{f-samp}(b)). Such surface and bulk modifications can be
patterned with high precision and their use for tailoring optical,
mechanical, and thermal properties are of interest. Optical
characterisation of such patterns at T-ray region was carried out.
Sapphire can be used as a substrate for optical elements due to
its high transparency (Fig.~\ref{f-sapp}) and performance of laser
inscribed optical elements in T-ray region has to be well
understood.

The TO phonon modes in $\alpha$-Al$_2$O$_3$ are at
385-388~cm$^{-1}$ (TO$_1$), 439~cm$^{-1}$ (TO$_2$) and
483~cm$^{-1}$ (LO$_2$) branches, 569~cm$^{-1}$ (TO$_3$) and
630~cm$^{-1}$ (LO$_3$), 633~cm$^{-1}$ (TO$_4$) and 1021~cm$^{-1}$
(LO$_4$)~\cite{book}; all the values were taken from Table 5.2 of
Ref.~\cite{book}. However, the best numerical fit of experimental
sapphire's reflectivity was achieved for LO$_4$ at 906~cm$^{-1}$
(not shown here). Figure~\ref{f-ripples}(a) shows a strong
alteration of reflectivity from laser ablated pattern of ripples
at the TO$_1$ mode at two different polariser orientations.
Self-organised ripple structures with period of $\sim 250$~nm
(Fig.~\ref{f-ripples}(a)) or $\Lambda_r \simeq \lambda_l/(2n)$
were recorded at the used laser wavelength $\lambda_l = 1030$~nm
and refractive index of sapphire $n = 1.7$. Strong relative
reflectivity increase at the position of LO$_2$ can be partially
caused by a very low value of $R$ for the c-plane sapphire sample
(a small values of denominator used in normalisation). There is an
observable change in $R$ at the spectral location of TO$_4$ mode.
Reflection from the laser ablated surface shows spectrally broader
TO$_{1,4}$ bands (Fig.~\ref{f-ripples}(a)), a sign of mode
softening. Reflectivity interrogated at different linear polariser
angles showed a consistent change in $R$ values of the TO$_{1,4}$
bands.

Figure~\ref{f-ref} shows transmission and reflection spectra from
the pattern of laser inscribed lines with period $\Lambda \simeq
24~\mu$m. In the reflected spectrum a broad band around $\lambda_R
= 294$~cm$^{-1}$  was observed (or $\lambda_R = 34~\mu$m in free
space with refractive index $n =1$). The reflected wavelength is
also the same which is effectively absorbed, hence, coupled into
the structure/sample. The momentum conservation for the
wavevectors of reflected light $\mathbf{k}_R$, structure/sample
mode $\mathbf{k}_s$ (could be a surface wave, SPP, TO-mode), and a
grating $\mathbf{k}_g$ is given by $\mathbf{k}_R = \mathbf{k}_s -
\mathbf{k}_g$. This is the simplest two wave mixing scenario
mediated by a grating. One can find the corresponding wavelength
$\lambda_s$ from the definition $k_s \equiv 2\pi/(\lambda_s/n)$,
at which the energy is deposited (absorbed) from the right
triangle rule:
\begin{equation}\label{e2}
\left(\frac{2\pi}{\lambda_R/n}\right)^2 =
\left(\frac{2\pi}{\lambda_s/n}\right)^2 -
\left(\frac{2\pi}{\Lambda}\right)^2.
\end{equation}
\noindent Equation~\ref{e2} allows to calculate the wavelength and
energy of the mode which is efficiently absorbing an incoming IR
light and, hence, effectively reflecting it. One would find
$\lambda_s = 30.94~\mu$m (323~cm$^{-1}$) with $n = 3.2$ in
sapphire for the high transmission region (Fig.~\ref{f-sapp}).
This did not fit the TO$_1$ mode. The second reflectivity peak at
366~cm$^{-1}$ (or $\lambda_R = 27.3~\mu$m in free space)
corresponds to the $\lambda_s = 25.64~\mu$m or 390~cm$^{-1}$
(Eqn.~\ref{e2}), which match the TO$_1$~\cite{book}. Since a
square lattice was inscribed in sapphire (Fig.~\ref{f-sapp}(c)),
there is a mode with period $\sqrt{2}\Lambda$. For this period the
reflection peak at 366~cm$^{-1}$ would correspond to $\lambda_s =
26.43~\mu$m or 378~cm$^{-1}$, which is also close to the TO$_1$.
For the opaque region (shaded in Fig.~\ref{f-samp}(a)), $n \simeq
(4.5-5.5)$ for the TO$_{1,2}$ region and the same analysis using
Eqn.~\ref{e2} suggest the 366~cm$^{-1}$ peak being close to the
TO$_1$ mode; similarly, the broad 294~cm$^{-1}$ feature has no
match with phonon mode. The broad 294~cm$^{-1}$ reflection band
can be due to a coupling with acoustic of hybrid
modes~\cite{Ishitani} and would require a systematic study for
different $\Lambda$ values matching the phonon modes.

Even though the reflectivity changes are very weak in the
normalised spectrum (Fig.~\ref{f-ref}), they are discernable on a
Rayleigh scattering slope (approximated by a line in this narrow
spectral window). This proves that a designed periodic pattern can
be used to couple energy into specific phonon mode.

\section{Conclusions and outlook}

It is demonstrated that surface and volume structuring of sapphire
has direct impact on softening of TO phonon modes and changing
reflectivity at a specific mode. Laser induced damage and
structural disorder is the reason of this modification of phonon
spectrum.  Control of phonon spectrum is strongly required for the
growth of layered structures where different optical and acoustic
modes are coupling with the interface waves~\cite{Ishitani}. Among
other possible applications are surface micro-texturing for ELO
solutions in GaN-based LEDs, an electron mobility control in
structures with strong phonon scattering, and in thermo-electrical
materials for enhancement of difference between electron and
phonon mean free path. Also a fast growing field of applications
is in the design of flat optical elements using surface
patterning~\cite{Nivas} where fs-laser surface and bulk
structuring has been well
developed~\cite{03apl2901,02ass705p,16l}.

\vspace{0.1cm} \small{This study was carried out as a part of the
Australian synchrotron beamtime proposal M8468-2014. Partial
support via Research Council Discovery grant DP130101205 and
collaboration project with Workshop of Photonics, Ltd. is highly
appreciated. Polished sapphire samples were donated by Tecdia Ltd.
Financial support from the Research Council of Lithuania under the
KITKAS project, contract No.LAT-04/2016. We acknowledge
Teravil-Ekspla for the access to a commercial T-Spec THz-TDS
system.}

\bibliographystyle{spiebib}
\small

\end{document}